\documentclass[10pt, conference, letterpaper]{ IEEEtran}
\IEEEoverridecommandlockouts
\usepackage{cite}
\usepackage{amsmath,amssymb,amsfonts}
\usepackage{graphicx}
\usepackage{textcomp}
\usepackage{xcolor}

\usepackage{amssymb}
\usepackage{amsmath}

\usepackage{algorithm}
\usepackage{algpseudocode}

\newcommand{\hl}[1]{{\textcolor{black}{#1}}}

\def\BibTeX{{\rm B\kern-.05em{\sc i\kern-.025em b}\kern-.08em
    T\kern-.1667em\lower.7ex\hbox{E}\kern-.125emX}}

\begin{document}

\title{FedLECC: Cluster- and Loss-Guided Client Selection for Federated Learning under Non-IID Data\\
}

\author{
\IEEEauthorblockN{
Daniel M. Jimenez-Gutierrez\textsuperscript{1},
Giovanni Giunta\textsuperscript{1},
Mehrdad Hassanzadeh\textsuperscript{1},
Aris Anagnostopoulos\textsuperscript{1},\\
Ioannis Chatzigiannakis\textsuperscript{1},
Andrea Vitaletti\textsuperscript{1}
}
\IEEEauthorblockA{
\textsuperscript{1}Sapienza University of Rome 
Via Ariosto 25
Rome, 00185, Italy \\
jimenezgutierrez@diag.uniroma1.it, giovanni.giunta@corner.ch, hassanzadeh.1961575@studenti.uniroma1.it, \\ aris@diag.uniroma1.it, ichatz@diag.uniroma1.it, vitaletti@diag.uniroma1.it
}
}

\maketitle

\begin{abstract}
Federated Learning (FL) enables distributed Artificial Intelligence (AI) across cloud–edge environments by allowing collaborative model training without centralizing data. In cross-device deployments, FL systems face strict communication and participation constraints, as well as strong non-independent and identically distributed (non-IID) data that degrades convergence and model quality. Since only a subset of devices (a.k.a clients) can participate per training round, intelligent client selection becomes a key systems challenge. This paper proposes \emph{FedLECC} (Federated Learning with Enhanced Cluster Choice), a lightweight, cluster-aware and loss-guided client selection strategy for cross-device FL. FedLECC groups clients by label-distribution similarity and prioritizes clusters and clients with higher local loss, enabling the selection of a small yet informative and diverse set of clients. Experimental results under severe label skew show that FedLECC improves test accuracy by up to 12\%, while reducing communication rounds by approximately 22\% and overall communication overhead by up to 50\% compared to strong baselines. These results demonstrate that informed client selection improves the efficiency and scalability of FL workloads in cloud–edge systems.
\end{abstract}


\section{Introduction}
\label{sec:intro}

The proliferation of Internet of Things (IoT) and edge devices has enabled a new class of distributed Artificial Intelligence (AI) applications spanning cloud–edge environments~\cite{soori2023internet}. These applications include predictive maintenance~\cite{arafat2024machine} and anomaly detection~\cite{wang2024exploring}
, where data is generated at the network edge and must be processed under strict latency, bandwidth, and privacy constraints. Centralizing such data in the cloud is often impractical due to communication overhead and regulatory concerns.

Federated learning (FL) has emerged as a key enabler for distributed AI across cloud–edge infrastructures, allowing collaborative model training without moving raw data off-device~\cite{mcmahan2017communication}. In practice, however, FL systems operate under significant systems and networking constraints: only a subset of edge devices, i.e.~clients, can participate in each training round due to limited bandwidth, energy budgets, straggler effects, and heterogeneous device capabilities. As a result, intelligent client selection becomes a central challenge in deploying FL at scale.

A major complication in cross-device FL is non-independent and identically distributed (non-IID) data, also known as data or statistical heterogeneity, which causes client updates to diverge and may slow convergence or degrade model quality~\cite{jimenez2023application}. Among different forms of non-IID data, label skew - where clients hold disjoint or highly imbalanced label distributions - has been shown to be particularly harmful, leading to strong client drift and unstable aggregation~\cite{jimenez2025thorough,aggregators_FL}. For this reason, this work focuses on label-based non-IID data. \hl{Among the different forms of non-IID data, label skew is widely recognized as one of the most challenging scenarios in FL, as it induces strong client drift and unstable aggregation. Such conditions naturally arise in cloud-edge deployments, where clients capture localized events or user-specific behaviors. In these settings, naive client selection strategies may waste communication resources on redundant or low-impact updates, motivating the need for intelligent selection mechanisms tailored to heterogeneous data distributions.}

Beyond non-IID data considerations, these challenges highlight a broader cloud–edge selection problem: how to select a small set of clients that provides both informative and diverse updates while respecting communication and participation constraints. Prior work has shown that uniform random sampling is often suboptimal in such environments~\cite{nagalapatti2022your}, motivating the need for intelligent, system-aware client selection mechanisms.

Personalized FL (PFL) further emphasizes the role of client selection under non-IID data. Tan et al.~\cite{tan2022towards} distinguish between regularization-based methods and selection-based strategies, the latter explicitly controlling which clients participate in training. Our work falls into this selection-based category, with a focus on scalable selection for cross-device FL systems.

In this paper, we propose \emph{FedLECC} (Federated Learning with Enhanced Cluster Choice), a lightweight, cluster-aware, loss-guided client selection strategy for cloud–edge FL. FedLECC groups clients based on similarity, via Hellinger distance (HD)~\cite{hd_properties}, in label distributions and prioritizes clusters and clients associated with higher local loss, i.e., edge devices where the current global model underperforms. By jointly enforcing diversity, through clustering, and informativeness (through loss-guided selection), FedLECC improves learning efficiency while significantly reducing communication overhead in highly non-IID settings.

\subsection{Motivation}
\hl{To address these challenges in a systematic manner, we next formalize the motivation and research questions that guide the design of FedLECC.} Deploying FL as a cloud–edge service requires balancing learning performance with system-level efficiency~\cite{kairouz2021advances}. Under non-IID data, local updates are biased toward client-specific distributions, which can slow convergence and lead to suboptimal global models~\cite{zhao2018federated}. These effects are amplified in large-scale cross-device deployments, where only a fraction of clients can be selected per round. Consequently, effective client selection is essential to avoid wasting communication resources on redundant or low-impact updates.

This work investigates whether combining clustering and loss-based client selection can improve both model performance and communication efficiency for federated workloads in cloud–edge environments, addressing the following research questions:

\begin{enumerate}
    \item \textbf{RQ1:} To what extent can FedLECC improve test accuracy under severe label skew in cross-device FL?
    \item \textbf{RQ2:} How do clustering-based diversity control and loss-guided prioritization contribute to FedLECC’s performance gains?
    \item \textbf{RQ3:} How much can FedLECC reduce communication rounds and total communication overhead compared to state-of-the-art client selection strategies?
\end{enumerate}

\subsection{Contributions}
The main contributions of this paper are:

\begin{enumerate}
    \item We propose FedLECC, an intelligent, cluster-aware client selection strategy for FL in cloud–edge environments where clients observe highly non-IID data.
        
    \item We show that by selecting a very limited but suitably selected number of informative edge devices can lead to significant improvements of learning efficiency while also drastically reducing communication costs.
    
    \item Through extensive experiments under severe label skew, we demonstrate that FedLECC improves accuracy by up to 12\% while reducing communication rounds by about 22\% and overall communication overhead by up to 50\% compared to strong baselines.

\end{enumerate}

\section{Related Work}
\label{sec:related}
\label{subsec:related-clustering}

Non-IID data, particularly \emph{label skew}, is a major challenge in FL, leading to slow convergence and degraded model accuracy when aggregating client updates~\cite{aggregators_FL,zhao2018federated}. In PFL, two main classes of approaches address non-IID data: \emph{regularization-based} methods, which modify local objectives, and \emph{selection-based} methods, which prioritize specific clients during training~\cite{tan2022towards}.

\subsection{Regularization-based Solutions}

Several works mitigate non-IID data by regularizing local training objectives. FedProx~\cite{li2020federated} extends FedAvg by introducing a proximal term that constrains local updates, improving robustness under non-IID data. FedNova~\cite{wang2021novel} normalizes local updates to account for heterogeneous computation and varying numbers of local steps, reducing bias introduced by stragglers. FedDyn~\cite{acar2021federated} introduces a dynamic regularization term that corrects client drift over time, achieving improved convergence under non-IID data.

While effective, these approaches do not explicitly address which clients should participate in each training round, nor do they exploit data diversity through structured client selection.

\subsection{Selection-based Solutions}

Client selection has emerged as a complementary strategy to address non-IID data and communication constraints. HACCS~\cite{wolfrath2022haccs} clusters clients based on label histograms and selects latency-efficient clients from each cluster, improving robustness to stragglers. FedCLS~\cite{li2022fedcls} leverages group label information and Hamming distance to guide client selection, achieving faster convergence than random sampling. FedCor~\cite{tang2022fedcor} models inter-client correlations using Gaussian Processes and selects clients based on their predicted contribution to global accuracy.

Loss-based selection has also proven effective. Cho et al.~\cite{pmlr-v151-jee-cho22a} showed that prioritizing clients with higher local loss accelerates convergence and improves accuracy, leading to the Power-of-Choice (POC) strategy.

Our work differs from prior selection-based approaches by \emph{jointly combining clustering and loss-guided selection}. FedLECC first structures the client population by label-distribution similarity and then prioritizes high-loss participants within each group, jointly promoting informative updates and diversity across suitably selected clients. This enables more targeted client participation, improving learning performance and significantly reducing communication cost under severe label skew.

\section{Problem Formulation}
\label{sec:problem}

We follow the standard FL setting introduced in prior work~\cite{pmlr-v151-jee-cho22a}. Consider a system with $K$ clients, where each client $i$ holds a local dataset $\mathcal{B}_i$ of size $N_i$, and datasets are disjoint across clients. Let $N=\sum_{i=1}^K N_i$ denote the total number of data samples.

Each client trains a local model parameterized by $\boldsymbol{\theta_i}$. Given a sample $\xi$, the per-sample loss is denoted by $f(\boldsymbol{\theta},\xi)$. The \emph{local empirical loss} of client $i$ is defined as:
\begin{equation}
\ell_i(\theta_i)= \frac{1}{N_i}\sum_{\xi \in \mathcal{B}_i} f(\boldsymbol{\theta_i}, \xi).
\end{equation}

\begin{figure*}[t]
    \centering
    \includegraphics[width=1.0\textwidth]{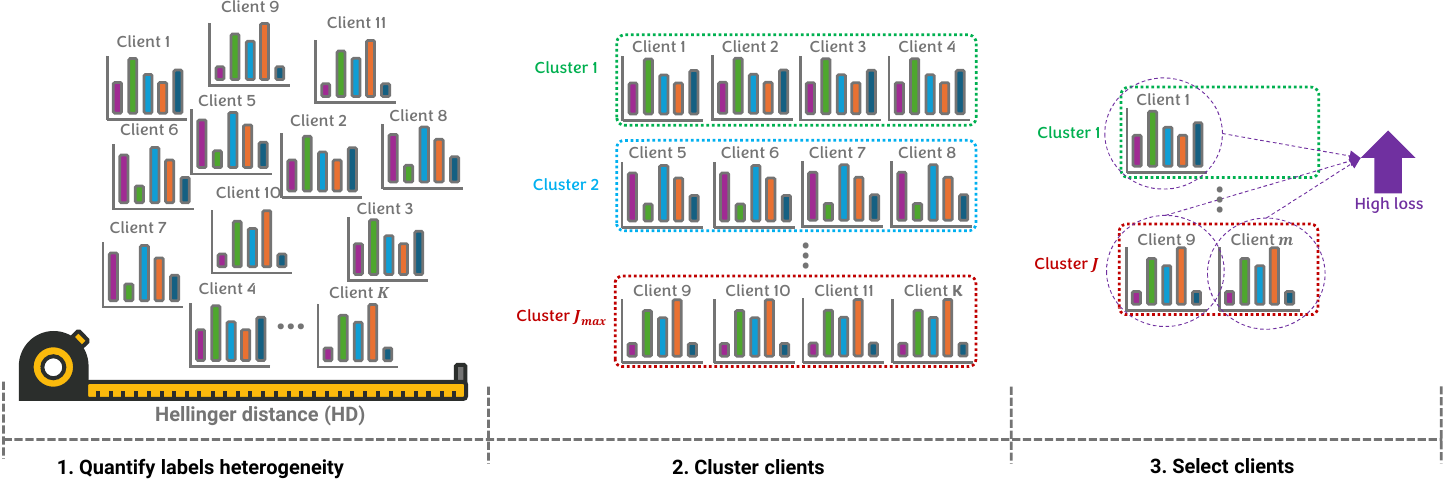}
    \caption{High-level client selection process in FedLECC.}
    \label{fig:high_level_process_FedLECC}
\end{figure*}

The objective of FL is to learn a global model $\boldsymbol{\theta}$ that minimizes the aggregated empirical risk:
\begin{equation}
\ell(\boldsymbol{\theta}) = \sum_{i=1}^{K} p_i \ell_i(\boldsymbol{\theta_i}),
\label{eq:problem_def}
\end{equation}
where $p_i = N_i/N$ reflects the relative data contribution of client $i$.

\subsection{Biased Client Selection}

FL proceeds in communication rounds coordinated by a central server. At each round, only a subset of $m \ll K$ clients is selected to participate based on the energy and latency constraints of the clients while aiming to maintain learning accuracy and reduce communication cost. Rather than selecting clients uniformly at random, \emph{biased client selection} assigns higher participation probability to clients that are expected to provide more informative updates~\cite{pmlr-v151-jee-cho22a}. Such strategies are particularly important under non-IID data, where client updates may vary significantly in usefulness.

\subsection{Clustering-based Client Selection}

In this work, we consider a clustering-based client selection strategy tailored to label-skewed FL settings, since label-based non-IID data is known to be the most detrimental form of non-IID data in FL. Clients are first grouped according to similarity in their label distributions. At each round, a subset of clusters is selected, and clients within these clusters are sampled based on their local loss values. Concretely, the selection process consists of:
\begin{enumerate}
    \item Grouping clients into clusters based on label-distribution similarity.
    \item Prioritizing clusters associated with higher average loss.
    \item Selecting a fixed number of clients per chosen cluster.
\end{enumerate}

This formulation enables controlling both \emph{diversity}, via clustering, and \emph{informativeness}, via loss-based selection, providing the foundation for the proposed FedLECC strategy. Remark that although the clients share information regarding the observed label distributions it does not constitute a leak of private information if existing, practical, privacy-preserving techniques such as Differential Privacy~\cite{erlingsson2014rappor} or secure multiparty computation~\cite{bohler2021secure} are integrated into the client selection pipeline.

\section{Proposed FedLECC Strategy}
\label{sec:cluster-strategy}
\begin{figure*}[t]
    \centering
    \includegraphics[width=1.0\textwidth]{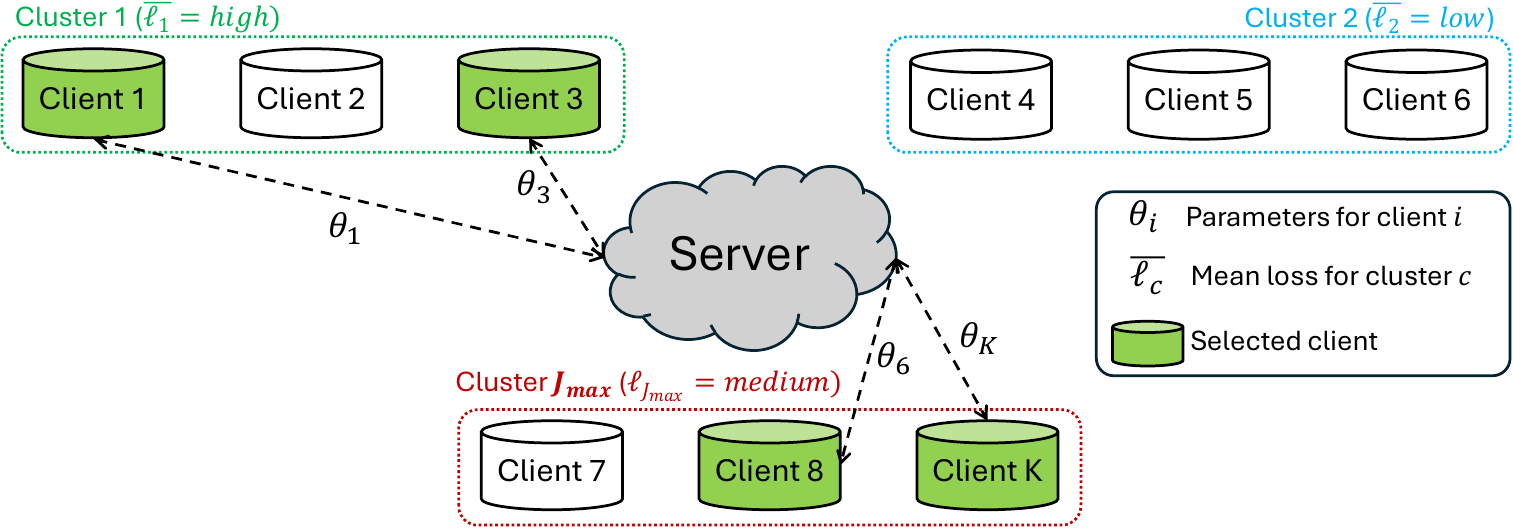}
    \caption{FL architecture with FedLECC client selection.}
    \label{fig:FL_architecture_FedLECC}
\end{figure*}

FedLECC is a client selection strategy designed for cross-device FL under severe label skew. As illustrated in Fig.~\ref{fig:high_level_process_FedLECC}, FedLECC operates in three stages:  
(1) Quantifying client non-IID data,  
(2) Clustering clients with similar label distributions, and  
(3) Selecting informative clients from representative clusters guided by the local empirical loss of each client.
This design explicitly balances \emph{diversity} and \emph{informativeness}, which are both critical for effective learning under non-IID data.

\hl{In practice, each client communicates its normalized label histogram to the server once, or whenever significant changes in the local data distribution occur. This information exchange is lightweight, as it scales with the number of labels rather than the dataset size, and can be amortized across many training rounds. No raw data or per-sample information is shared, preserving the decentralized nature of FL.}

\subsection{Non-IID data Quantification via Label Distributions}

FedLECC starts by characterizing non-IID data across clients using their label distributions. Each client $i$ provides the server with an aggregated histogram of labels, which does not reveal individual samples. The server computes pairwise distances between clients using the HD, a bounded and symmetric metric well suited for comparing probability distributions.

Alternative distance definitions based on feature distributions conditioned on labels were evaluated, but label-distribution distances consistently yielded better clustering quality and downstream performance. Recall that FedLECC focuses on label-based non-IID data, which is known to be the most detrimental form of non-IID data in FL.

\subsection{Clustering Clients}
\label{subsec:approach-clustering-clients}

Using the pairwise HD matrix, clients are grouped into clusters of similar label distributions. We evaluate several clustering techniques, including DBSCAN, $k$-medoids, and OPTICS. Among them, OPTICS consistently provides the best trade-off between cluster quality and robustness, as it does not require specifying the number of clusters in advance and adapts well to varying client densities.

Clustering plays a critical role in FedLECC by preventing the repeated selection of clients with highly similar data distributions, which could otherwise lead to the over-specialization of the global model. Purely loss-based client selection may repeatedly select clients from a single difficult data mode, leading to over-specialization. By enforcing cluster-level representation, FedLECC ensures that selected clients span diverse label data distributions, preserving diversity while still prioritizing challenging regions of the data space.

\subsection{Selecting Clusters and Clients}
\label{subsec:approach-selecting-clusters-clients}

At each communication round, FedLECC selects a subset of clusters and clients based on local loss values. Two parameters control the selection process:
\begin{itemize}
    \item $J$: Number of clusters selected ($J \le J_{\max}$),
    \item $m$: Total number of participating clients, with $z=\lceil m/J \rceil$ clients drawn per selected cluster.
\end{itemize}

Here, $J_{\max}$ is automatically determined by OPTICS based on the label distribution similarities among clients, as measured by the HD. After local training, each client reports its computed local empirical loss to the server. The server computes the average loss for each cluster and ranks clusters accordingly. The top-$J$ clusters are selected, and within each cluster, the $z$ clients with the highest loss are chosen. If a cluster contains fewer than $z$ clients, remaining slots are filled by high-loss clients from the next clusters, under the existing loss-based ordering.

\hl{To clarify the overall framework, FedLECC operates as a lightweight extension of the standard FL workflow. The server first acquires coarse-grained information about client non-IID data in the form of label histograms, which are used to compute inter-client similarities and derive clusters. During training, the clustering structure is treated as fixed, while client selection is performed dynamically at each communication round based on the reported local losses. Importantly, FedLECC does not modify the local training procedure or the aggregation rule, but only influences which clients participate in each round.}

The full selection procedure is summarized in Algorithm~\ref{alg:fedlecc-selection} and illustrated in Fig.~\ref{fig:FL_architecture_FedLECC}.

\begin{algorithm}[h]
\caption{Cluster- and Loss-Based Client Selection (FedLECC)}
\label{alg:fedlecc-selection}
\begin{algorithmic}[1]
\Require
Clusters $\mathcal{C}=\{C_1,\dots,C_{J_{\max}}\}$, target clusters $J$, target clients $m$
\Ensure Set $S$ of $m$ selected clients
\State $z \gets \lceil m / J \rceil$
\For{each client $i$}
    \State Compute local loss $\ell_i$ and send to server
\EndFor
\For{each cluster $C_k$}
    \State Compute mean loss $\bar{\ell}_k$
\EndFor
\State Select top-$J$ clusters with highest $\bar{\ell}_k$
\For{each selected cluster}
    \State Select top-$z$ clients with highest $\ell_i$
\EndFor
\If{$|S|<m$}
    \State Fill remaining slots with highest-loss clients from the following clusters ordered by descending $\bar{\ell}_k$
\EndIf
\State \Return $S$
\end{algorithmic}
\end{algorithm}



\hl{\subsection{Convergence Considerations.}
FedLECC modifies the standard FL workflow only through the client selection mechanism, while preserving the local training procedure and the server-side aggregation rule (e.g., weighted averaging as in FedAvg). As a result, FedLECC inherits the convergence properties of biased client selection schemes studied in prior work~\cite{pmlr-v151-jee-cho22a}. In particular, selecting clients with higher local loss increases the probability of sampling informative updates, which has been shown to accelerate convergence under non-IID data without destabilizing training.}

\hl{Moreover, the clustering step in FedLECC does not alter model updates, but enforces diversity among selected clients by preventing repeated sampling from highly similar data distributions. This mitigates client drift under severe label skew and empirically improves convergence stability, as observed in Section~V. While a formal convergence proof is beyond the scope of this work, our experimental results indicate that FedLECC converges faster and more stably than both uniform sampling and single-factor selection baselines.}

\hl{\subsection{Design Rationale and Implicit Optimization Objective}
FedLECC is guided by the objective of maximizing the utility of each communication round under a strict participation budget, i.e., selecting a small set of clients that is both informative and diverse under severe non-IID data. Loss-guided prioritization targets informativeness: clients (and clusters) with higher loss are more likely to yield updates that reduce the global objective, as also motivated by biased client selection studies}~\cite{pmlr-v151-jee-cho22a}\hl{. Clustering based on label-distribution similarity acts as a diversity control mechanism that prevents repeatedly sampling highly similar clients, mitigating over-specialization and improving robustness under label skew. Thus, the two-stage selection in Algorithm~\ref{alg:fedlecc-selection} can be seen as a practical approximation to constrained utility maximization that balances loss reduction with coverage of heterogeneous data distributions.}

\section{Experiments and Results}
\label{sec:experiments}
This section evaluates FedLECC in terms of (i) test accuracy and (ii) communication overhead under severe label skew. We compare against FedAvg and state-of-the-art baselines.

\subsection{Experimental Setup}

\textbf{Datasets and non-IID partitioning.}
We evaluate on MNIST~\cite{deng2012mnist} and FMNIST~\cite{xiao2017fashion} (see Table~\ref{tab:characteristics_datasets}). Each dataset is partitioned across $K$ clients using FedArtML~\cite{jimenez2024fedartml} with a Dirichlet($\alpha$) label split to simulate label skew. We focus on a high non-IID data regime (HD $\approx 0.9$), where larger values indicate highly non-IID data.

\begin{table}[t]
    \centering
    \caption{Characteristics of the datasets used in our evaluation.}
    \resizebox{1.0\columnwidth}{!}{
    \begin{tabular}{||c|c|c|c|c||}
    \hline
    Dataset & \#train & \#test & \#features & \#classes \\
    \hline\hline
    MNIST  & 60{,}000 & 10{,}000 & 784 & 10 \\
    FMNIST & 60{,}000 & 10{,}000 & 784 & 10 \\
    \hline
    \end{tabular}}
    \label{tab:characteristics_datasets}
\end{table}

\textbf{Model and training protocol.}
For both datasets, we train a Multilayer Perceptron (MLP) with two hidden layers (200 neurons), using cross-entropy loss and Stochastic Gradient Descent (SGD) with a learning rate of $0.005$ and a batch size of $64$. We use $T=150$ communication rounds, and report results averaged over five random seeds.


\textbf{Baselines and tuning.}
All baselines are tuned following their original papers: FedProx~\cite{li2020federated}, FedNova~\cite{wang2021novel}, FedDyn~\cite{acar2021federated}, HACCS~\cite{wolfrath2022haccs}, FedCLS~\cite{li2022fedcls}, FedCor~\cite{tang2022fedcor}, and POC~\cite{pmlr-v151-jee-cho22a}.

\subsection{Accuracy Comparison}

\begin{figure}[t]
    \centering
    \includegraphics[width=1\columnwidth]{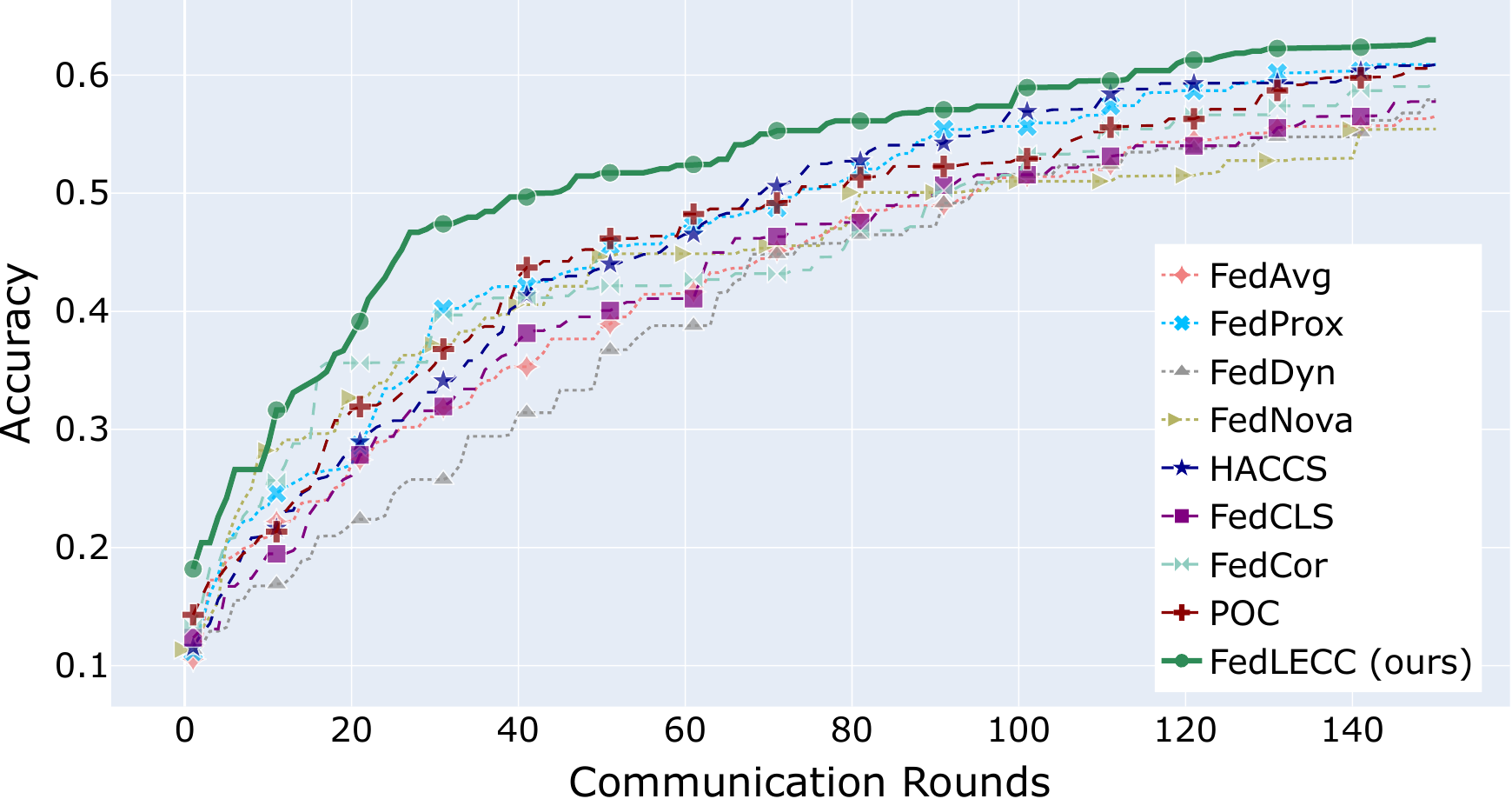}
    \caption{Accuracy comparison for the baselines and FedLECC using FMNIST and $K=100$ (best tuned hyperparameters).}
    \label{fig:clustered_accuracy}
\end{figure}

Figure~\ref{fig:clustered_accuracy} reports representative learning curves on FMNIST with $K=100$ clients under severe label skew. FedLECC converges faster than all baselines and reaches a higher final accuracy, indicating that its cluster-aware and loss-guided selection enables more effective use of each communication round. In this setting, FedLECC reduces the number of communication rounds required to reach a given accuracy level by approximately 22\% compared to FedAvg. In contrast, baselines relying on uniform sampling or single-factor selection exhibit slower convergence and larger performance fluctuations, which are symptomatic of client drift under non-IID data.

Table~\ref{tab:comparing_competitors_mnist_fmnist} reports final test accuracy on MNIST and FMNIST under highly non-IID data. FedLECC achieves the highest accuracy in most settings, especially with larger client populations, where non-IID data is more pronounced. Across these configurations, FedLECC improves model accuracy by up to 12\% compared to FedAvg and other strong baselines. Compared to both regularization-based and selection-based approaches (e.g., HACCS and POC), FedLECC’s combination of cluster-level diversity and loss-guided prioritization proves more effective than relying on either mechanism alone.

\begin{table}[t]
\centering
\caption{Accuracy (mean $\pm$ std) on MNIST and FMNIST under high non-IID data.}
\resizebox{\columnwidth}{!}{
\begin{tabular}{||c|cc|cc||}
\hline
& \multicolumn{2}{c|}{MNIST} & \multicolumn{2}{c||}{FMNIST} \\ \hline
Method & $K=100$ & $K=250$ & $K=100$ & $K=300$ \\ \hline\hline
HD & 0.90 & 0.86 & 0.90 & 0.86 \\
Silhouette & 0.641 & 0.502 & 0.723 & 0.409 \\ \hline
FedAvg  & 0.681 $\pm$ 0.02 & 0.745 $\pm$ 0.03 & 0.565 $\pm$ 0.03 & 0.634 $\pm$ 0.03 \\
FedProx & 0.681 $\pm$ 0.03 & 0.745 $\pm$ 0.03 & 0.608 $\pm$ 0.02 & 0.636 $\pm$ 0.02 \\
FedNova & 0.696 $\pm$ 0.03 & 0.732 $\pm$ 0.03 & 0.554 $\pm$ 0.03 & 0.598 $\pm$ 0.04 \\
FedDyn  & 0.654 $\pm$ 0.03 & 0.687 $\pm$ 0.02 & 0.567 $\pm$ 0.03 & 0.661 $\pm$ 0.03 \\
HACCS   & 0.625 $\pm$ 0.03 & 0.535 $\pm$ 0.03 & 0.608 $\pm$ 0.03 & 0.658 $\pm$ 0.03 \\
FedCLS  & 0.636 $\pm$ 0.02 & 0.644 $\pm$ 0.03 & 0.577 $\pm$ 0.03 & 0.639 $\pm$ 0.03 \\
FedCor  & 0.678 $\pm$ 0.03 & 0.546 $\pm$ 0.02 & 0.592 $\pm$ 0.03 & 0.652 $\pm$ 0.03 \\
POC     & 0.673 $\pm$ 0.03 & 0.719 $\pm$ 0.02 & 0.605 $\pm$ 0.03 & 0.668 $\pm$ 0.02 \\
\textbf{FedLECC (ours)} & \textbf{0.702 $\pm$ 0.03} & \textbf{0.772 $\pm$ 0.03} & \textbf{0.629 $\pm$ 0.03} & \textbf{0.675 $\pm$ 0.02} \\
\hline
\end{tabular}}
\label{tab:comparing_competitors_mnist_fmnist}
\end{table}

\subsection{Communication Overhead}

We measure the total communication exchanged between the server and clients over training, including model parameters, cluster information, and loss values. Table~\ref{tab:comm_overhead_mnist_fmnist} reports the average communication overhead under high non-IID conditions. FedLECC consistently achieves lower overhead than FedAvg by limiting participation to a small set of informative clients, and remains competitive with other selection-based baselines. Across the evaluated configurations, FedLECC reduces the overall communication overhead by up to 50\% compared to strong baselines, demonstrating that suitably selecting clients can significantly reduce bandwidth and coordination costs in cloud–edge FL systems without sacrificing accuracy.

\begin{table}[t]
\centering
\caption{Average communication overhead (MB) on MNIST and FMNIST (smaller is better).}
\resizebox{1.0\columnwidth}{!}{
\begin{tabular}{||c|cc|cc||}
\hline
& \multicolumn{2}{c|}{MNIST} & \multicolumn{2}{c||}{FMNIST} \\ \hline
Method & $K=100$ & $K=250$ & $K=100$ & $K=300$ \\ \hline\hline
FedAvg  & 49.28 & 123.21 & 49.28 & 126.28 \\
FedProx & 21.12 & 49.28  & 44.73 & 119.73 \\
FedNova & 17.70 & 132.63 & 73.61 & 93.92 \\
FedDyn  & 41.31 & 38.05  & 15.22 & 35.61 \\
HACCS   & 3.42  & 3.23   & 3.42  & 55.29 \\
FedCLS  & 35.26 & 40.21  & 35.20 & 114.30 \\
FedCor  & 2.47  & 4.47   & 3.97  & 41.80 \\
POC     & 4.78  & 5.65   & 2.52  & 34.31 \\
\textbf{FedLECC (ours)} & \textbf{2.11} & \textbf{1.93} & \textbf{2.24} & \textbf{33.55} \\
\hline
\end{tabular}}
\label{tab:comm_overhead_mnist_fmnist}
\end{table}




\section{Discussion}
\label{sec:discussion}
FedLECC consistently improves learning performance while reducing system-level costs, even when only a limited subset of clients is selected. This confirms that informed selection, rather than broad participation, is essential for scalable cross-device FL.

With respect to RQ1, FedLECC improves test accuracy under severe label skew by prioritizing informative and representative edge devices. Explicitly considering non-IID data during client selection helps mitigate client drift and improve the stability of the aggregation process. Regarding RQ2, FedLECC’s gains stem from combining cluster-based diversity control with loss-guided prioritization, which encourages the selection of clients with diverse label distributions while focusing training on underperforming regions of the data distribution.

RQ3 is addressed through the observed reductions in communication rounds and overall communication overhead. Selecting fewer yet more informative clients reduces bandwidth consumption and coordination costs, which are critical constraints in cloud–edge infrastructures. From a systems perspective, these results position FedLECC as an effective selection mechanism for resource-efficient distributed AI workloads.

Finally, from a scalability perspective, our experiments consider configurations with up to $K=300$ clients across different datasets and severe non-IID scenarios. The results show that FedLECC preserves its accuracy advantages over strong baselines as the number of participating clients increases, indicating good scalability.

Overall, the discussion highlights that FedLECC aligns with the goals of intelligent cloud computing and networking by addressing core challenges in selection and scalable FL under non-IID data.

\section{Potential Challenges for FedLECC}
\label{sec:challenges}

While effective, FedLECC’s performance depends on configuration choices such as the number of selected clusters and clients. Similar sensitivity to participation-related parameters is also present in other state-of-the-art client selection baselines, such as POC, FedCor and HACCS. Inappropriate configurations may trade off accuracy for communication efficiency or vice versa. Addressing this common challenge through adaptive parameter tuning and workload-aware configuration mechanisms remains an important direction for future work.

\section{Conclusion and Future Work}
\label{sec:conclusion}
This paper presented FedLECC, an intelligent, cluster-aware client selection strategy for FL workloads in cloud–edge environments under highly non-IID data and more specifically severe label skew. By combining cluster-level diversity control with loss-guided prioritization, FedLECC enables the server to select a small set of informative edge devices at each communication round. Experimental results under highly non-IID label skew show that FedLECC improves model accuracy by up to 12\%, while reducing communication rounds by approximately 22\% and overall communication overhead by up to 50\%. These gains demonstrate that informed client selection can substantially improve the efficiency and scalability of cross-device FL systems operating under limited bandwidth and participation budgets.

Several directions remain open for future work from a systems and networking perspective. In particular, exploring lightweight mechanisms to automatically adapt FedLECC’s configuration to workload dynamics and resource availability could further improve its robustness in cloud–edge environments. In addition, integrating privacy-preserving techniques such as Differential Privacy~\cite{erlingsson2014rappor} or secure multiparty computation~\cite{bohler2021secure} into the client selection pipeline remains an important direction to strengthen trust and security guarantees in practical federated cloud–edge systems.

\section{Acknowledgments}
Daniel M.\ Jimenez-Gutierrez was partially supported by PNRR351
TECHNOPOLE -- NEXT GEN EU Roma Technopole -- Digital Transition,
FP2 -- Energy transition and digital transition in urban regeneration and
construction. 
Aris Anagnostopoulos was supported by the PNRR
MUR project PE0000013-FAIR, the PNRR MUR project IR0000013-
SoBigData.it, and the MUR PRIN project 2022EKNE5K ``Learning in
Markets and Society.'' 
Ioannis Chatzigiannakis was supported by PE07-
SERICS (Security and Rights in the Cyberspace) -- European Union
Next-Generation-EU-PE00000014 (Piano Nazionale di Ripresa e Re-
silienza -- PNRR). 
Andrea Vitaletti was supported by the project SERICS
(PE00000014) under the MUR National Recovery and Resilience Plan
funded by the European Union - NextGenerationEU.

\bibliographystyle{IEEEtran}
\bibliography{references}

\end{document}